\documentclass[twocolumn,showpacs,aps,prl,amsmath,amssymb,floatfix]{revtex4}
\usepackage{graphicx}
\usepackage{dcolumn}
\usepackage{bm}
\begin{document}
\title{Critical disorder effects in Josephson-coupled
quasi-one-dimensional superconductors}
\author{E. Nakhmedov$^{1,2}$ and R. Oppermann$^{1,3}$}
\affiliation{ $^1$Institut f\"ur Theoretische Physik,
Universit\"at W\"urzburg,
D-97074 W\"urzburg, Germany\\
$^2$Institute of Physics, Azerbaijan National Academy of
Sciences,H. Cavid str. 33, AZ1143 Baku, Azerbaijan\\
$^3$Institut de Physique Th\'eorique, CEA Saclay, Orme des
Merisiers, F-91191 Gif-sur-Yvette, France}
\date{\today}
\begin{abstract}
Effects of non-magnetic randomness on the critical temperature
$T_c$ and diamagnetism are studied in a class of quasi-one
dimensional superconductors. The energy of Josephson-coupling
between wires is considered to be random, which is typical for
dirty organic superconductors. We show that this randomness
destroys phase coherence between the wires and $T_c$ vanishes
discontinuously when the randomness reaches a critical value. The
parallel and transverse components of the penetration depth are
found to diverge at different critical temperatures $T_c^{(1)}$
and $T_c$, which correspond to pair-breaking and phase-coherence
breaking. The interplay between disorder and quantum phase
fluctuations results in quantum critical behavior at $T=0$,
manifesting itself as a superconducting-normal metal phase
transition of first-order at a critical disorder strength.
\end{abstract}
\maketitle
Quasi-one-dimensional (quasi-1D) organic conductors, including the
charge-transfer (Bechgaard) salts of $(TMTSF)_2X$ (where $TMTSF$
stands for tetramethyltetraselenofulvalinium and $X = PF_6,
ClO_4,NO_3$ being a strong electron acceptor or anion) \cite{bj09}
and A-15 compounds \cite{wg73}, attract enhanced interest since
the discovery of superconductivity in $(TMTSF)_2PF_6$.
Low temperature properties of the organic superconductors are very
sensitive to disorder. Alloying anions, $x$-ray irradiation and
cooling rate controlled anion reorientation introduce non-magnetic
randomness into the system while leaving the backbone structure
and the unit cell of the organic superconductors to a large extent
unchanged. There is a common agreement that disorder, introduced
by means of these experimental methods, must be characterized as
non-magnetic, and yet it was shown \cite{cchh82, iys98, japm04} to
suppress the superconducting (SC) phase.

Effect of disorder on the SC phase has a long-standing history.
According to the Anderson's theorem \cite{anderson59}, the SC
critical temperature $T_c$ for $s$-wave pairing is insensitive to
the scattering rate on non-magnetic impurities. Magnetic
impurities break time-reversal symmetry of the $s$-pairing,
suppress at the same time the SC phase \cite{ag58}. Strong
disorder of non-magnetic impurities may however destroy $d$-wave
pairing \cite{ml85}. Interplay between superconductivity and
Anderson localization in a strongly disordered superconductor was
shown \cite{ml85, grt01,ro90, szh01, swt04, sok08, mrms08, gl01,
sf05,fink87} to result in spatial inhomogeneity of the order
parameter.
High purity of the organic superconductor
backbone even in the dirty limit seems to exclude a spatial
inhomogeneity of the order parameter modulus along SC wires,
offering an opportunity for another mechanism of disorder-driven
superconductor-normal metal phase transition. Effects of order
parameter phase fluctuations on $T_c$ have also been studied in
low-dimensional superconductors \cite{el74,fy76,ek95,nf98}. It is
well known that there is no SC phase transition in 1D and
two-dimensional (2D) systems \cite{rice65}, since strong
fluctuations of the order parameter phase destroy off-diagonal
long-range order (ODLRO) in a single SC wire or film. Strong phase
fluctuations in clean quasi-1D superconductors have been shown
\cite{el74,fy76} to suppress $T_c$ below a mean-field transition
temperature. Classifying the superconductors with small stiffness
as bad metals, Emery and Kivelson have evaluated \cite{ek95} a
critical temperature of phase ordering by formally dividing a
clean bulk superconductor into small regions with well defined
phase, and have shown strong suppression of SC phase by phase
fluctuations. Nevertheless, effects of disorder on phase
fluctuations are neglected in all of these papers.

In contrast to these previous activities we study in this Letter a
suppression of superconductivity as a result of the destruction of
the order parameter phase coherence by disorder. We consider
weakly linked quasi-one-dimensional superconductors with random
Josephson-couplings between pure one-dimensional (1D) SC wires.
Singlet pairing is considered to occur inside the wires.
Therefore, we assume that non-magnetic randomness does not affect
the order parameter amplitude. We demonstrate in this Letter that
(i) non-magnetic randomness in the Josephson-coupling destroys
correlation of the phases between different chains in quasi-1D
superconductors even in the classical phase fluctuation regime,
(ii) randomness yields quantum critical behavior. A SC-normal
metal phase transition occurs at $T=0$ with increasing the
strength of disorder, and that (iii) a suppression of the SC phase
occurs discontinuously in both classical and quantum phase
fluctuations as a first-order phase transition when the
disorder-strength reaches a critical value. We derive that
parallel and perpendicular components of the penetration depth,
$\lambda_{\|}$ and $\lambda_{\perp}$ diverge at different critical
temperatures $T_c^{(1)}$ and $T_c$, which correspond to
pair-breaking in the wires and to phase coherence breaking between
the SC wires, respectively.

{\em Classical fluctuations of the phase.} The free energy
functional of a quasi-1D superconductor weakly linked with
Josephson coupling energy $E_{\bf j, j+g}$ between
nearest-neighbor chains can be written in the presence of the
magnetic field ${\bf B}$ as
\begin{eqnarray}
&&F_{st} = N_s^{(1)}(T) \xi_{\|} \sum_{\bf j} \int dz \bigg
\{\frac{\hbar ^2}{8 m_{\|} \xi_{\|}^2} \bigg( \frac{\partial
\varphi_{\bf j}}{\partial z} - \frac{2e \xi_{\|}}{\hbar c}
A_z\bigg)^2 +
\nonumber\\
&&\sum_{{\bf g} = \pm 1} E_{\bf j,j+g} [1 - \cos \big(
\varphi_{\bf j} - \varphi_{\bf j+g} + \frac{2e \xi_{\|}}{\hbar
c}\int_{\bf j}^{\bf j+g} {\bf A}_{\perp} d {\bf r}_{\perp} \big)]
+
{}\nonumber\\
&& + \xi_{\|} a^2_{\perp} \frac{({\bf B}({\bf r}) - {\bf
B}_{ext})^2}{8 \pi} \bigg \}, \label{freeenergy}
\end{eqnarray}
where $\varphi_{\bf j}(z)$ denotes the order parameter phase,
${\bf A} = \{{\bf A}_{\perp}, A_z \}$ is the vector-potential, and
$N_s^{(1)}(T)= N_s^{(1)}(0) \tau(T)$ is the linear density of SC
electrons with $\tau (T) = \frac{T_c^{(1)} - T}{T_c^{(1)}}$ and
$N_s^{(1)}(0) \equiv N_N^{(1)} \simeq \frac{p_F}{\hbar}$ at $T \le
T_c^{(1)}$. Dimensionless coordinates ${\bf r} = \{ {\bf j}, z \}$
are introduced on the scale of longitudinal $\xi_{\|} =
\frac{\hbar^2 N_s^{(1)}(0)}{4 m_{\|} T_c^{(1)}}$ and transverse
$\xi_{\perp} \sim a_{\perp}$ components of the coherence length.
We assume the Josephson energy $E_{\bf j, j+g}$ to be a random
parameter with Gaussian distribution given by
\begin{equation}
P \{ E_{\bf j, j+g} \} = \frac{1}{\sqrt{2 \pi W^2}} \exp \big \{ -
\frac{(E_{\bf j, j+g} - E_{\bf g})^2}{2 W^2} \big \}.
\label{gauss}
\end{equation}
Employing the replica trick one can integrate out the Gaussian
disorder to obtain the average value of the free energy $
\mathcal{F} =-T \langle \ln Z \rangle$ as
\begin{eqnarray}
\mathcal{F} = -T \int \prod_{\bf j, g}
\frac{N_s^{(1)}\xi_{\|}}{\sqrt {2 \pi}}d \zeta_{\bf j, g} e^{-
\frac{N_s^{(1) 2}\xi_{\|}^2}{2}\zeta_{\bf j, g}^2} \times \nonumber\\
 \times \ln \int \prod \mathcal{D} \varphi_{\bf j}
e^{-F/T}\hspace{+3.2cm}
\label{eff-avenergy}\\
{\rm with}\quad F = N_s^{(1)}\xi_{\|} \sum_{\bf j} \int dz \bigg\{
\frac{\hbar^2}{8 m_{\|} \xi_{\|}^2} \left( \frac{\partial
\varphi_{\bf j}}{\partial z} \right)^2 +
\nonumber\\
\hspace{-2.6cm} \sum_{\bf g} (E_{\bf g}-
N_s^{(1)}\xi_{\|}W\zeta_{\bf j,g}) [1 - \cos(\varphi_{\bf j}
-\varphi_{\bf j+g})] \bigg\}.
\label{eff-energy}
\end{eqnarray}
where $\zeta_{\bf j, g}$ is a Hubbard-Stratonovich auxiliary
field. The average value of a given functional $C(\{\varphi_{\bf
j} \})$, e.g. $\cos \varphi_{\bf j}$ or $\cos(\varphi_{\bf
j}-\varphi_{\bf j+g})$, can be obtained according to the relation
$\langle\langle C(\{\varphi_{\bf j} \})\rangle\rangle
=-T\frac{\delta}{\delta \eta_{\bf j}}\langle \ln Z\rangle
|_{\eta_{\bf j}=0}$ by adding the source term $\sum_{\bf j}\int dz
\eta_{\bf j} C(\{\varphi_{\bf j} \})$ to the free energy
functional, which yields for the correlator
\begin{eqnarray}
&&\langle \langle C(\{ \varphi_{\bf j} \})\rangle \rangle = \int
\prod_{\bf j, g} \frac{N_s^{(1)}\xi_{\|}}{\sqrt {2 \pi}}d
\zeta_{\bf j, g} e^{- \frac{N_s^{(1) 2}\xi_{\|}^2}{2}\zeta_{\bf j,
g}^2} \times \nonumber\\
&&\times \frac{\int \mathcal{D} \varphi C (\{\varphi_{\bf j}\})
e^{- F/T}}{\int \mathcal{D} \varphi e^{- F/T}}, \label{av}
\end{eqnarray}
where the double bracket $\langle \langle \dots \rangle \rangle$
means averaging over thermodynamic fluctuations and over
randomness. In order to estimate an asymptotic behavior of the
correlator, e.g. $\langle \langle \cos(\varphi_{\bf j} -
\varphi_{\bf j+g}) \rangle \rangle$ we write the integrand of
Eq.(\ref{av}) as $\exp \{-N_s^{(1) 2}\xi_{\|}^2 f(\zeta_{\bf
j,g})\}$, and apply the stationary-phase approximation to
determine an extremal value of the auxiliary field $\bar
{\zeta}_{\bf j,g}$ minimizing the function $f(\zeta_{\bf j,g})$.
The saddle point value of $\zeta_{\bf j,g}$ is obtained to be
$\bar{ \zeta}_{\bf j, g} =\frac{W}{T}\int dz
\big(\langle\cos(\varphi_{\bf j}(z)-\varphi_{\bf j+g}(z))\rangle -
\frac{ \langle \cos(\varphi_{\bf j}(z)-\varphi_{\bf j+g}(z))
\cos(\varphi_{\bf j}(0)-\varphi_{\bf j+g}(0))\rangle}{\langle
\cos(\varphi_{\bf j}(0)-\varphi_{\bf j+g}(0))\rangle} \big)$. The
constant $N_s^{(1)}\xi_{\|}$ on the exponent can be estimated to
be equal to $N_s^{(1)}\xi_{\|}\simeq \frac{\epsilon_F}{T_c^{(1)}}
\sim 10^3$ for the organic superconductors with $\epsilon_F$ being
the Fermi energy, which ensures a sharply peaked saddle point of
the integrand. The thermodynamic averages in the expression of
$\bar{\zeta}_{\bf j,g}$ are taken with the free energy functional,
given by Eq.(\ref{eff-energy}) at the saddle point $\zeta_{\bf j,
g}=\bar{\zeta}_{\bf j, g}$. So, a contribution of the non-magnetic
randomness to the effective free-energy functional is proportional
to the variance of the phase correlator, which gives an idea on
the form of the disorder-dependent term in the effective
functional.

The critical temperature for the quasi-1D superconductors can now
be found from Eq.(\ref{av}),written for $\cos\varphi_{\bf j}$ by
using the self-consistent mean-field method \cite{el74}, which
consists in replacing the phase correlations of the cosine term by
$\sum_{\bf g} E_{\bf g} [1-\cos(\varphi_{\bf j}(z)-\varphi_{\bf
j+g}(z))] \to E_{\perp} [1 - \langle \langle \cos(\varphi) \rangle
\rangle_{eff} \cos (\varphi (z))]$, where $E_{\perp}  = \sum_{\bf
g} E_{\bf g}$. For a clean system $\langle \langle \cos(\varphi)
\rangle \rangle_{eff}$ was chosen \cite{el74} to be equal to
$\langle \cos(\varphi) \rangle$. For the disordered superconductor
we choose $\langle\langle \cos \varphi \rangle\rangle_{eff}=
\langle\langle \cos \varphi \rangle\rangle - N_s^{(1)}\xi_{\|}
\frac{\langle\langle \cos \varphi \rangle^2\rangle-\langle\langle
\cos \varphi \rangle\rangle^2}{\langle\langle \cos \varphi
\rangle\rangle}$. Such a form of $\langle\langle \cos \varphi
\rangle\rangle_{eff}$ is similar to the expression of the saddle
point value for the averaged order parameter. Taking advantage of
the smallness of $(E_{\perp}- N_s^{(1)}\xi_{\|} W \zeta)
\langle\langle\cos(\varphi)\rangle\rangle_{eff}$ near $T_c$, we
expand both the numerator and the denominator of the integrand of
Eq.(\ref{av}), written for $\langle \langle \cos(\varphi) \rangle
\rangle_{eff}$, in this parameter. The thermodynamic averages
become pure one-dimensional after this expansion, which can be
taken easily, yielding a power series of $\zeta$ for the
integrand. Therefore, the integration over $\zeta$ is immediately
performed. Since all higher order in $\langle \langle
\cos(\varphi) \rangle \rangle_{eff}$ terms of the expansion vanish
at $T=T_c$, we get the equation for $T_{c}$
\begin{eqnarray}
\hspace{-2cm} 1 &=& \frac{E_{\perp} N_s^{(1)}
\xi_{\|}}{T_c}\biggl(1 - \frac{W^2
\xi_{\|} N_s^{(1)} \eta ^2}{ T_c E_{\perp}}\biggr)\times\nonumber\\
&&\hspace{1cm}\times \int \langle \cos(\varphi(0))
\cos(\varphi(z))\rangle dz, \label{Tc}
\end{eqnarray}
where $\eta$ is the coordination number. The phase correlator in
Eq.(\ref{Tc}) is calculated in the clean limit of the $1D$ free
energy functional, obtained from Eq.(\ref{freeenergy}) by setting
$E_{\bf j,j+g} = 0$, which returns (see Ref.\cite{rice65})
\begin{equation}
\langle \cos (\varphi (0)) \cos(\varphi (z)) \rangle = \exp
\{-|z|/r_c \},
\label{1Dcorrelator}
\end{equation}
where $r_c = \hbar^2 N_s^{(1)}(T)/2 m_{\|} \xi_{\|}T$. Using
dimensionless $T_c$-shift $t = \sqrt{\eta \epsilon_F E_{\perp}}
\left(\frac{1}{T_c} - \frac{1}{T_{c}^{(1)}} \right)$ and
disorder parameter $q = \frac{W^2}{E_{\perp}} \sqrt{\frac{2 m_{\|}
\xi_{\|}^2 \eta}{\hbar^2 E_{\perp}}} = \frac{ W^2}{2 E_{\perp}
T_c^{(1)}} \sqrt{\frac{\eta \epsilon_F}{E_{\perp}}}$,
Eqs.(\ref{Tc},\ref{1Dcorrelator}) yield
\begin{equation}
1 = t^2 (1 -q t).
\label{Tc3}
\end{equation}
Expanding the physical solution of this cubic equation in the weak
disorder regime (small $q$) the $T_c$-shift obeys
\begin{equation}
\frac{1}{T_c} = \frac{1}{T_c^{(1)}}+ \frac{1}{\sqrt{\eta
\epsilon_F E_{\perp}}} + \frac{1}{T_c^{(1)}} \bigg(\frac{W}{2
E_{\perp}} \bigg)^2,
\label{criticalT}
\end{equation}
showing that $T_c$ decreases with increasing randomness like
$W^{2}$. For a pure system Eq.(\ref{criticalT}) gives the
dependence  $T_c \sim E_{\perp}^{1/2}$, in agreement with Efetov
and Larkin in Ref. \cite{el74}. This expression shows that even a
small interchain-coupling sets up an ODLRO in the system, and
consequently, the critical temperature increases with $E_{\perp}$.
On the other hand, disorder reduces $T_c$ due to "melting" of the
order parameter phase coherence between neighboring chains.
\begin{figure}
\resizebox{.48\textwidth}{!}{%
\includegraphics[width=1cm]{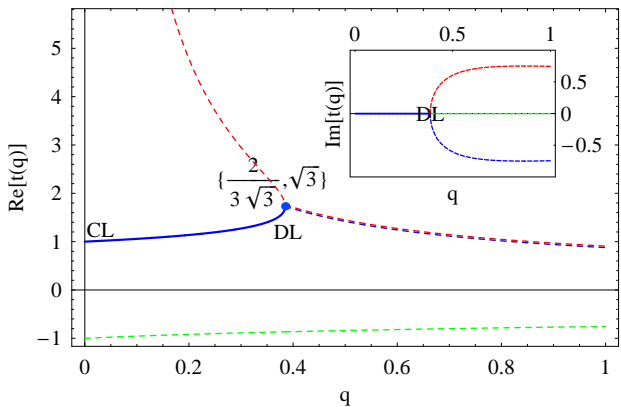}}
\caption{The physical solution $t(q)$, giving the
$T_c(W)$-dependence, within the full range from clean limit (CL:
$q=0$) to the dirty limit (DL: $q_c=2/3\sqrt{3})$) is highlighted
as the bold (blue) curve. Formal solutions of the cubic
Eq.(\ref{Tc3}) are shown for completeness. $T_c(q)$ vanishes
abruptly at $q=q_c$.}
\label{Tc-classic}
\end{figure}
The exact solution of Eq.(\ref{Tc3}) gives three roots, among
which the physical one is confined to the finite $q$-range as
shown in Fig.\ref{Tc-classic} by the bold line. According to this
solution, the critical temperature decreases monotonically with
increasing $q$ in the interval of $0 \le q \le q_c =\frac{2}{3
\sqrt 3}$. The SC phase hence suppressed beyond the critical
disorder-value $W_c^2 =\frac{4 E_{\perp} T_c^{(1)}}{3}
\sqrt{\frac{E_{\perp}}{3 \eta \epsilon_F}}$, being transformed
into a normal metallic phase for $W^2 > W_c^2$. The critical
temperature drops to zero at $W^2 = W_c^2$ with a jump of size
$\Delta T_c = T_c^{\ast}= \left( \sqrt{\frac{3}{\eta \epsilon_F
E_{\perp}}} + \frac{1}{T_c^{(1)}}\right)^{-1}$. Thus the SC-normal
metal phase transition appears as a 1st-order transition.

In order to find the behavior of $t$ near the disorder limit (DL
in Fig.\ref{Tc-classic}) $\{t^{\ast}, q^{\ast}\} = \{\sqrt{3},
\frac{2}{3 \sqrt{3}} \}$, we expand $\delta t = t^{\ast} - t$ in
terms of $\delta q = q^{\ast} - q$, which gives
$\delta t = 3^{3/4}\sqrt{\delta q}$. In other words, $\delta T_c =
T_c - T_c^{\ast}$ behaves as $\delta T_c = \frac{3^{3/4} T_c
T_c^{\ast}}{E_{\perp}(4 \eta \epsilon_F
E_{\perp}(T_c^{(1)})^2)^{1/4}} (W_c^2 - W^2)^{1/2}$ in the
vicinity of the breakdown point $\{ T_c^{\ast}, W_c^2 \} = \bigg\{
\left( \sqrt{\frac{3}{\eta \epsilon_F E_{\perp}}} +
\frac{1}{T_c^{(1)}}\right)^{-1}, \frac{4 E_{\perp} T_c^{(1)}}{3}
\sqrt{\frac{E_{\perp}}{3 \eta \epsilon_F}} \bigg \}$.

{\em Quantum phase fluctuations regime.}
We shall improve the calculation of the phase-correlators by
taking into account the transverse rigidity of the system, which
provides a more realistic determination of the transition
temperature in the quantum fluctuation regime. We start from the
Lagrangian, for simplicity at ${\bf B}=0$
\begin{equation}
\mathcal{L} = \frac{K \xi_{\|}(0)}{8} \sum_{\bf j} \int dz [\hbar
\dot{\varphi}_{\bf j}(z)]^2 - F_{eff}\{\varphi \}
\label{lagrangian}
\end{equation}
where $\dot{\varphi}$ denotes the time derivative of the phase.
The dynamical term in $\mathcal{L}$ can be interpreted as the
electrostatic energy of charged wires \cite{fy76, nf98}
$E_{el}=\frac{1}{2} \sum_{\bf i,j} \int dz \int dz' C_{\bf
i,j}(z-z')V_{\bf i}(z)V_{\bf j}(z')$ produced according to the
first Josephson equation $\dot{\varphi} = (2e/\hbar)V$; $C_{\bf
i,j}(z-z')$ are the specific coefficients of electrostatic
induction. Rewriting $E_{el}$ in terms of time-derivative of
phases, the Fourier transform $K({\bf q_{\perp}}, q_z)$ of  the
new coefficients $K_{\bf i,j}(z-z')=\frac{1}{4e^2}C_{\bf
i,j}(z-z')$, has the physical meaning of a compressibility.  In
Eq.(\ref{lagrangian}) we neglect a spatial dispersion of the
compressibility and take $K({\bf q_{\perp}},q_z) = K = $ const.
$F_{eff}\{\varphi \}$ is the functional $F$ in
Eq.(\ref{eff-energy}), written at the saddle point
$\tilde{\zeta}_{\bf j,g}$ of the averaged free energy
$\mathcal{F}$. The saddle point of $ \mathcal{F}$ is found to be
$\tilde{\zeta}_{\bf j,g} = \frac{W}{\mathcal{F}_{eff}} \int dz
\langle \langle [1-\cos(\varphi_{\bf j}(z)-\varphi_{\bf j+g}(z))]
\rangle \rangle$, where $\mathcal{F}_{eff}=-T \ln \int \prod_{\bf
j,g}\mathcal{D} \varphi_{\bf j} e^{-F_{eff}/T}$. So,
$\mathcal{F}_{eff}$ has to be calculated self-consistently.
Note that the model would be calculated more rigorously by
replacing the 1D wire with discrete analogue of Josephson-coupled
cells, and considering a strongly anisotropic 3D Josephson
network. After averaging over disorder, one can introduce the
'order parameters' as $\zeta_a = \langle \langle e^{i\varphi_{\bf
j}^a}\rangle \rangle$ and $q_{a,b}^{\pm} = \langle \langle e^{i
\varphi_{\bf j}^a}\rangle \langle e^{\pm i \varphi_{\bf
j'}^b}\rangle \rangle$, where $\zeta_a$ and $q_{a,b}$ are the
order parameters corresponding to the SC and glassy phases with
$a, b$ being the replica indices. The model can be mapped to the
solvable Sherrington-Kirkpatrick model for the long-ranged
phase-phase correlations, nevertheless a solution of the model for
the short-ranged (nearest-neighbor) phase-phase correlation case,
realized in our model, is hard task.


The Hamiltonian, expressed in terms of the phases $\varphi_{\bf
j}$ and canonical conjugate momenta $\Pi_{\bf j}$ as
$\mathcal{H}=\sum_{\bf j} \hbar \int \Pi_{\bf j}\dot{\varphi}_{\bf
j} dz - \mathcal{L}$, becomes
\begin{eqnarray}
\mathcal{H} = \sum_{\bf j} \int dz \bigg \{ 2 \frac{ \Pi_{\bf
j}^2(z)}{K \xi_{\|}(0)} + \frac{\hbar^2 N_s^{(1)}(T)}{8m_{\|}
\xi_{\|}}\biggl[ \bigg( \frac{\partial \varphi_{\bf j}}{\partial
z} \bigg)^2 +{}
\nonumber\\
{} + \sum_{\bf g} \delta_{cl}^2 [1 - \cos(\varphi_{\bf j}(z) -
\varphi_{\bf j + g}(z))]\biggr] \bigg\},
\label{hamilton}
\end{eqnarray}
where $\Pi_{\bf j} = \frac{1}{\hbar}\frac{\delta {\mathcal
L}}{\delta \dot{\varphi}_{\bf j}}=\frac{1}{4}\hbar K \xi_{\|}(0)
\dot{\varphi}_{\bf j}$, and $\delta_{cl}$ as given by
\begin{equation}
\delta_{cl}^2 = \delta_{0}^2 \biggl[ 1 - \frac{W^2 N_s^{(1)}
\xi_{\|}}{E_{\perp} \mathcal{F}_{eff}} \langle \langle [1 - \cos
(\varphi_{\bf j}(z) - \varphi_{\bf j+g}(z))]\rangle \rangle
\biggr] \label{rigidity-dis}
\end{equation}
represents either the dimensionless anisotropy-parameter or the
transverse rigidity of the random system, while $\delta_{0}$ in
Eq.(\ref{rigidity-dis}) being the transverse rigidity of the pure
system
\begin{equation}
\delta_{0} = \left(\frac{E_{\perp}}{\hbar^2/8 m_{\|} \xi_{\|}^2}
\right)^{1/2} = \frac{(\epsilon_F E_{\perp})^{1/2}}{T_c^{(1)}}.
\label{rigidity-pure}
\end{equation}
$\delta_0$ is a natural small parameter of the quasi-1D
superconductors, which ensures small interchain-coupling energies
in comparison with the intrachain Cooper-pair energy. Indeed,
expressing $E_{\perp}$ through the interchain tunneling integral
$J_{\perp}$ \cite{el74} as $E_{\perp} \simeq
J_{\perp}^2/\epsilon_F$ yields $\delta_0 =J_{\perp}/T_c^{(1)}$.

The quantum description is realized by expressing $\varphi_{\bf
q}$ and $\Pi_{\bf q}$ as a linear superposition of Bose operators
$b_{\bf q}$
as $\varphi_{\bf q} = \big ( \frac{\pi \alpha \bar{\omega}}{\omega
({\bf q})} \big )^{1/2} (b_{\bf q} + b_{- \bf q}^{\dag})$ and
$\Pi_{\bf q} = i \big( \frac{\omega ({\bf q})}{4 \pi \alpha
\bar{\omega}} \big)^{1/2} (b_{- \bf q} - b_{\bf q}^{\dag})$. In
the framework of the self-consistent harmonic approximation
(SCHA), we rewrite the expansion of the cosine operator in
Eq.(\ref{hamilton}) in terms of the bosonic particle number
operator ${\hat N}_{\bf q}=b_{\bf q}^{\dag} b_{\bf q}$ \cite{fy76,
nf98} as
\begin{equation}
\hat{\mathcal H} = \sum_{\bf q} \hbar \omega ({\bf q}, T) \left (
b_{\bf q}^{\dag} b_{\bf q} + 1/2 \right ),
\label{hamiltonian}
\end{equation}
where the eigenfrequency of oscillation $\omega ({\bf q}, T)$ is
given
\begin{equation}
\omega ({\bf q},T) = \bar {\omega} [q_z^2 + \delta _{cl}^2 e^{-
S_{\alpha}^{(0)}({\bf g}, T)} 2 (2- \cos q_x - \cos q_y)]^{1/2}.
\label{frequency}
\end{equation}
We express the amplitude of the frequency
$\bar{\omega}=(N_s^{(1)}(T)/m_{\|} K \xi_{\|}^2)^{1/2}$ as
$\bar{\omega} = 2 \pi \alpha T_c^{(1)}\tau^{1/2}/\hbar$, where
\begin{equation}
\alpha = 2 \big(m_{\|} / K \hbar^2 N_s^{(1)} \big)^{1/2}/\pi.
\label{alpha}
\end{equation}
The parameter $\alpha$, which is an essential parameter of the
theory, can assume values between zero and one \cite{el74}.

The factor $\exp \{- S_{\alpha}({\bf g}, T)\}$ in
Eq.(\ref{frequency}) is obtained as
\begin{equation}
S_{\alpha}({\bf g},T) = \frac{2 \pi \alpha \bar{\omega}}{N}
\sum_{\bf q} \frac{1 - \cos ({\bf q}_{\perp} {\bf g})}{\omega
({\bf q}, T)}\left({N}_{\bf q} + \frac{1}{2} \right), \label{S}
\end{equation}
where $N$ is the number of unit cells per volume, and ${N}_{\bf q}
= \{ \exp \left(\hbar \omega ({\bf q}, T)/T \right) - 1\}^{-1}$ is
Planck's distribution function for phonons. A physical meaning of
$\exp \{- S_{\alpha}({\bf g}, T)\}$ is an average of
$\cos(\varphi_{\bf j}(z) - \varphi_{\bf j+g}(z))$ over all
one-phonon states,  $\exp \{- S_{\alpha}({\bf g}, T)\} =\langle
\langle \cos (\varphi_{\bf j}(z) - \varphi_{\bf j+g}(z))\rangle
\rangle^{(T)}$, \cite{fy76}. Application of the SCHA results in a
{\em renormalization} of the parameter $\delta_{cl}$, changing
thus the oscillation frequency $\omega ({\bf q},T)$ by means of
the phase-phase correlator
\begin{equation}
\delta_{cl}^2 \to \delta_{qu}^2(T) = \delta_{cl}^2 \exp\{-
S_{\alpha}({\bf g},T) \}.
\label{delta-quT}
\end{equation}
Note that the SCHA is valid under the condition $\sum_{\bf
q}|A_{\bf q}|^2\langle N_{\bf q} \rangle =\frac{2 \pi \alpha
\bar{\omega}}{N} \sum_{\bf q} \frac{1 - \cos ({\bf q}_{\perp} {\bf
g})}{\omega ({\bf q}, T)}\langle{N}_{\bf q}\rangle <1$, which
means that few phonons are excited in the system.

It is easier to see that $\mathcal{F}_{eff}$ can be calculated as
$\mathcal{F}_{eff}= - T \ln Tr \{ e^{-\hat{\mathcal H}/T} \}$ by
neglecting the dynamical term ($K=0$ or $\alpha=0$) in Eq.(\ref
{hamilton}), which gives $\mathcal{F}_{eff} = T$.

\begin{figure}
\hspace{-.5cm}\resizebox{.5\textwidth}{!}{%
\includegraphics[width=1cm]{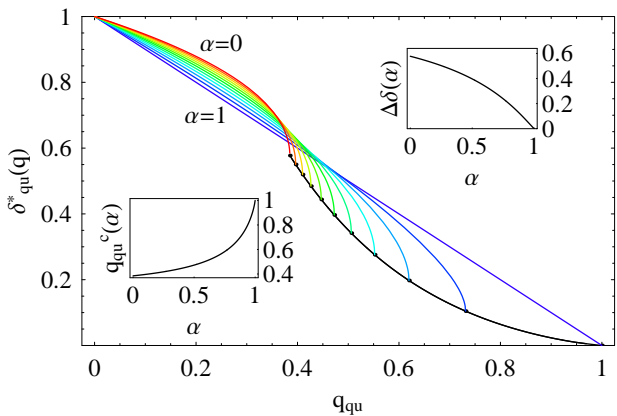}}
\caption{The dependence of the reduced $T=0$ transverse rigidity
$\delta_{qu}^{\ast}(q_{qu})$ on the disorder-strength parameter
$q_{qu}$ is shown for $0\leq\alpha\leq 1$ in steps of
$\Delta\alpha=0.1$. At $q_{qu}=q_{qu}^c$,
$\delta_{qu}^{\ast}(q_{qu})$ drops to zero for $\alpha <1$ and
vanishes continuously only at $\alpha =1$. Inserts show the
$\alpha$-variation of the jump (upper right corner) and of its
position $q_{qu}^c(\alpha)$ (lower left).}
\label{Tc-quantum}
\end{figure}

Let us start with the {\em $T=0$ limit}: expressing the
phase-phase correlator $e^{-S_{\alpha}({\bf g},0)}$ in terms of
$S_{\alpha}({\bf g},0) =\frac{\pi \alpha \bar{\omega}}{N}
\sum_{\bf q} \frac{1 - \cos ({\bf q}_{\perp} {\bf g})}{\omega
({\bf q}, 0)}$ gives $e^{-S_{\alpha}({\bf g},0)} =
(\delta_{qu}(0))^{\alpha}\equiv \delta_{qu}^{\alpha}$, which
implies that even small interchain-coupling stabilizes ODLRO,
hence also a finite $T$ phase transition should exist. In order to
get an explicit expression for the dependence of $ \delta_{qu}$ on
$\delta_0$ and on disorder, we have to solve the equation
$\delta_{qu}^2 = \delta_{cl}^2 e^{-S_{\alpha}({\bf g}, 0)}$
together with Eq.(\ref{rigidity-dis}) for $\delta_{cl}$, the
latter of which also depends on $\delta_{qu}$. Thus the equation
for the reduced transverse rigidity $\delta_{qu}^{\ast} =
\delta_{qu}/\delta_{qu}^{(0)}$, where $\delta_{qu}^{(0)} =
\delta_0^{\frac{2}{2 - \alpha}}$ is the renormalized transverse
rigidity for the clean system at $T = 0$, assumes the form
\begin{equation}
(\delta_{qu}^{\ast})^{3 - 2\alpha} = (\delta_{qu}^{\ast})^{1 -
\alpha} - q_{qu},
\label{delta-qu}
\end{equation}
where the quantum parameter of randomness $q_{qu}$ reads
\begin{equation}
q_{qu} = \frac{C W^2}{2 E_{\perp}^2} \delta_0^{\frac{2}{2 -
\alpha}}.
\label{qofW}
\end{equation}
The numerical solution of Eq.(\ref{delta-qu}) is depicted in
Fig.\ref{Tc-quantum}. The reduced $T=0$ transverse rigidity
$\delta_{qu}^{\ast}(q_{qu})|_{T=0}$ is shown to decrease with
increasing disorder for (fixed) $\alpha < 1$, and suddenly drops
to zero at the critical disorder strength $q_{qu} = q_{qu}^{c}$.
Hence the quantum critical behavior corresponds to a first order
phase transition. Fig.\ref{Tc-quantum} shows how the breakdown
point shifts with increasing $\alpha$ to larger disorder, and the
jump vanishes for $\alpha\rightarrow 1$.
Eq.(\ref{delta-qu}) becomes linear for $\alpha = 1$ and gives, by
inferring the $q_{qu}(W)$-relation from Eq.(\ref{qofW}),
\begin{equation}
\delta_{qu}(W)|_{T=0,\alpha=1} = \delta_0^2 \bigg[ 1 - \frac{C W^2
\epsilon_F}{2 E_{\perp} T_c^{(1) 2}}\bigg].
\end{equation}
Here, the transverse rigidity $\delta_{qu}(W)|_{T=0,\alpha=1}$
decreases linearly with increasing $W^2$ and vanishes at $W_c^2 =
\frac{2 E_{\perp} \left(T_c^{(1)}\right)^2}{C \epsilon_F}$. The
quantum critical behavior in the model is however controlled by
two parameters, the strength of randomness $q_{qu}(W)$ and the
parameter of quantum dynamics $\alpha$. For $\alpha < 1$, the
superconductor-normal metal phase transition at $T=0$ is always
discontinuous, and only turns into second-order at $\alpha = 1$.

Let us now study the {\it finite $T$ behavior} of the transverse
rigidity. The phase transition in a quasi-1D superconductor occurs
at some temperature $T = T_c$ when the transverse rigidity in the
ensemble of phases $ \{ \varphi_{\bf j}(z) \}$ vanishes. The
energy spectrum $\omega ({\bf q}_{\perp}, q_z)$ of the collective
excitations is reorganized and the transverse ${\bf
q}_{\perp}$-dependent part of $\omega ({\bf q}_{\perp}, q_z)$
vanishes at $T = T_c$, i.e. symmetry breaking occurs in the
bosonic excitation at $T = T_c$. Inserting the solution of
Eq.(\ref{S}) for $T < \alpha T_c^{(1)}$ into $\delta_{qu}^2(T) =
\delta_{cl}^2 e^{- S_{\alpha}({\bf g}, T)}$ and using
Eq.(\ref{delta-quT}), we obtain
\begin{equation}
\delta_{qu}^2(T) = \delta_{qu}^2(0)\left(\frac{T}{\alpha
T_{c0}}\right)^{\alpha} \exp\bigg \{-C
\frac{T}{T_{c0}}\frac{\delta_{qu}(0)}{\delta_{qu}(T)}\bigg \},
\label{deltaTC}
\end{equation}
where a new temperature scale is introduced by means of $T_{c0} =
\delta_{qu}(0) T_c^{(1)}$, and $C$ is a constant $C \sim 1$. In
terms of $y =\left(\frac{\alpha T_{c0}}{T}\right)^{\alpha /2}
\frac{\delta_{qu}(T)}{\delta_{qu}(0)}$ and $\theta = \left(
\frac{T}{T_{c0}} \right)^{1 -
\frac{\alpha}{2}}\frac{C}{2}\alpha^{\alpha/2}$, Eq.(\ref{deltaTC})
assumes the form $y = exp \{ - \theta /y \}$, which has a non-zero
solution only for $\theta \leq e^{-1}$. The finite solution of
this equation vanishes discontinuously at $\theta = \theta_c=
e^{-1}$, giving the following value for $T_c$
\begin{equation}
T_c = T_{c0}{\alpha}^{- \frac{\alpha}{2 - \alpha}} (2/e
C)^{\frac{2}{2 - \alpha}}.
\end{equation}
The magnitude of the jump in $y(\theta_c)$ is $e^{-1}$, and hence
the phase transition is of first-order.

{\em Meissner effect.} The current density is calculated according
to $\frac{1}{c}{\bf J}(z,{\bf j}) = - T\frac{\delta}{\delta {\bf
A}} \langle \ln Z({\bf A})\rangle $.
For simplicity we present here only the diamagnetic contribution
to the $i$-th ($i = \|, \perp $) component of the current
\begin{equation}
J_i^{dia}(z,{\bf j})= - \frac{c}{4\pi \lambda_i^2} A_i(z,{\bf j}),
\end{equation}
where the longitudinal- $\lambda_{\|}$ and the transverse
$\lambda_{\perp}$ components of the penetration depth are obtained
as
\begin{equation}
\lambda_{\|}^{-2} = \frac{4 \pi e^2 N_s^{(1)}}{c^2 m_{\|}
a_{\perp}^2}; \quad \frac{\lambda_{\|}^2}{\lambda_{\perp}^2} =
\frac{2 m_{\|}a_{\perp}^2 E_{\perp}}{\hbar^2}\langle \langle \cos
(\varphi_{\bf j} - \varphi_{\bf j+g}) \rangle \rangle.
\end{equation}
Although $\lambda_{\|}(T)$ diverges at $T=T_c^{(1)}$ due to pair
breaking in the SC wires, $\lambda_{\perp}(T)$ diverges at the
global SC transition temperature $T=T_c$, where the phase
coherence between neighboring wires is destroyed. Randomness in
the Josephson coupling shifts $T_c$ to lower temperatures and,
therefore, the magnetic field parallel to the SC wires penetrates
easier into the organic superconductor. On the other hand, the
randomness does not break the Cooper pairs, keeping the
penetration of a perpendicular magnetic field into the SC wires
unchanged.

In this Letter we studied disorder-effects on $T_c$ and on the
diamagnetism of Josephson-coupled quasi-1D superconductors.
Interplay of disorder with quantum phase fluctuations plays a
central role for superconductor-normal metal phase transitions in
quasi-1D superconductors. The quantum criticality is controlled by
two quantities, namely disorder strength and dynamical parameter
of phase fluctuations. The present model's quantum criticality
signals the existence of a quantum critical phase between SC- and
normal phase. Its nature, whether it is a "mixing" of a glassy and
CDW or SDW phases, needs further investigation.

We thank the DFG for support under grant Op28/7-1.

\end{document}